# Machine Learning-Assisted Exploration of Thermally Conductive Polymers Based on High-Throughput Molecular Dynamics Simulations

Ruimin Ma[1], Hanfeng Zhang,[1] Jiaxin Xu,[1] Yoshihiro Hayashi[2], Ryo Yoshida[2], Junichiro Shiomi[3], Tengfei Luo[1*]

1. Department of Aerospace and Mechanical Engineering, University of Notre Dame, Notre Dame, Indiana 46556, United States

2. The Institute of Statistical Mathematics, Research Organization of Information and Systems, Tachikawa, Tokyo, 190-8562, Japan

3. Department of Mechanical Engineering, University of Tokyo, Bunkyo-ku, Tokyo, 113-8656, Japan

* Corresponding email: tluo@nd.edu

## ABSTRACT

Finding amorphous polymers with higher thermal conductivity is important, as they are ubiquitous in heat transfer applications. With recent progress in material informatics, machine learning approaches have been increasingly adopted for finding or designing materials with desired properties. However, relatively limited effort has been put on finding thermally conductive polymers using machine learning, mainly due to the lack of polymer thermal conductivity databases with reasonable data volume. In this work, we combine high-throughput molecular dynamics (MD) simulations and machine learning to explore polymers with relatively high thermal conductivity (> 0.300 W/m-K). We first randomly select 365 polymers from the existing PolyInfo database and calculate their thermal conductivity using MD simulations. The data are then employed to train a machine learning regression model to quantify the structure-thermal conductivity relation, which is further leveraged to screen polymer candidates in the PolyInfo database with thermal conductivity > 0.300 W/m-K. 133 polymers with MD-calculated thermal conductivity above this threshold are eventually identified. Polymers with a wide range of thermal conductivity value are selected for re-calculation under different simulation conditions, and those polymers found with thermal conductivity above 0.300 W/m-K are mostly calculated to maintain values above this threshold despite fluctuation in the exact values. A classification model is also constructed, and similar results were obtained compared to the regression model in predicting polymers with thermal conductivity above or below 0.300 W/m-K. The strategy and results from this work may contribute to automating the design of polymers with high thermal conductivity.

**INTRODUCTION**

Bulk amorphous polymers are usually regarded as thermal insulators despite that they are ubiquitous in heat transfer applications, such as thermal interface materials, heat exchangers, and thermal energy storage mediums. [1-4] To improve thermal conductivity (TC) of bulk polymers, efforts have focused on compositing thermally conductive fillers with amorphous polymer matrices.[5-7] Different fillers, such as metals,[8, 9] ceramics,[10, 11] and carbon,[12, 13] have been intensively investigated to enhance the TC of polymer composites. However, it has been pointed out that the polymer matrix TC has a major impact on the composite TC, with higher matrix TC leading to greater enhancement in composite TC.[1, 14] As a result, finding amorphous polymers that intrinsically have high TC will be of great importance to their heat transfer applications.

Despite the progress in understanding thermal transport physics in polymers,[15-20] relatively limited information is known which type of amorphous polymer is better in conducting heat, and experimentally screening a large number of natural and synthesized polymers is very challenging. The current largest polymer database, PoLyInfo [21], has a very limited number of pure polymers with TC values, and among them, many have spread values from different sources. It would be ideal if a structure-TC relation can be established for quickly screening a large number of polymers to identify promising candidates for further experimental exploration. Recently, data-driven approaches leveraging machine learning techniques have been developed to establish structure-property relations for a variety of materials.[22-26] However, such effort for polymer TC is very limited up to date. Wu et al. leveraged a technique called transfer learning[27] to establish a structure-TC surrogate model for polymers with a predictive accuracy of $R^2$=0.732 based on only 28 experimental TC data from PolyInfo.[28] They were able to design a polymer with the TC of 0.410 W/m-K using this surrogate model, which is notably high for amorphous polymers. However, while the transfer learning technique helped to solve the small data issue,[28] it is always favorable that big data are available for machine learning studies, which can potentially increase the accuracy of the surrogate models for a large diversity of polymers.

In this work, we leveraged high-throughput molecular dynamics (MD) simulations to generate TC data for quantifying the structure-TC relation in polymers. 365 polymers from the PoLyInfo database were randomly selected for MD simulations, with 324 (89%) of them having MD-labeled TC below 0.300 W/m-K. Random forests (RF) models were employed to learn the structure-TC relation on those 365 data thereafter, which was in turn used to screen polymer candidates with TC > 0.300 W/m-K in the whole PoLyInfo database. 104 polymer candidates with RF-predicted TC exceeding 0.300 W/m-K were then labeled using MD simulations, with 92 (88%) of them having MD-labeled TC above this threshold. While it is found that the MD-calculated TC can vary from one simulation condition to another due to the stochastic nature of amorphous polymer morphology, those predicted to have TC > 0.300 W/m-K mostly remained to be above this threshold in different simulation conditions. A classification model was also constructed, and similar results were obtained compared to the regression model in predicting polymers with TC above or below 0.300 W/m-K. Our work may provide useful guidance and an integral component to the experimental exploration of high TC polymers.

**METHODS**

The general procedure of the TC data generation using high-throughput MD simulation is depicted in **Figure 1a**. It consists of two main steps including amorphous structure optimization and non-equilibrium MD (NEMD) TC calculation as detailed below.

**Amorphous polymer structure generation.** The structure of 12,777 homopolymers were manually collected from the PoLyInfo database,[21] which were stored in their monomer format and represented using the SMILES[29] (Simplified Molecular Input Line Entry System). We built a Python pipeline based on PYSIMM[30] to automatically generate amorphous polymer structures taking the SMILES strings of monomers as the inputs. The pipeline first polymerized the monomer into a polymer chain. The number of atoms per chain was controlled to be around 600 for all simulated polymers, and the resultant chain length of all the simulated polymers ranged from ~6 to ~40 nm. The variation in chain length is due to the fact that some polymers have branches, but some others do not. In the meantime, PYSIMM assigned the GaFF2 (General AMBER Force Field 2) forcefield[31] parameters to the polymers, which also generates input scripts for MD simulations using the large-scale atomic-molecular massively parallel simulator (LAMMPS).[32] Polymers with missing GAFF2 forcefield parameters were skipped in the data generation process. In all simulations, periodic boundary conditions in all spatial directions were used. Each polymer chain was then duplicated by 6 copies and enclosed in a simulation box. The structure was optimized via several steps. In the first few steps, the polymer system was simulated with the electrostatic interactions turned off, and for the Lennard-Jones interactions, a cutoff of 0.300 nm was used. The system was first simulated under the NVT (constant number of atoms, pressure, and temperature) ensemble at 100 K for 2ps, with timestep 0.1 fs applied. After that, the system was heated up from 100 K to 1000 K in 1 ns under the NVT ensemble and then simulated under the NPT (constant number of atoms, pressure, and temperature) ensemble for 50 ps at 0.1 atm and 1000 K. This step was conducted to further relax the structure and ensure the elimination of close contacts. The system was further simulated under NPT at 1000 K for 1 ns, where the pressure was allowed to increase from 0.1 atm to 500 atm, with a timestep of 1 fs and SHAKE constraints[33] applied. The use of SHAKE constraint allows larger timestep of 1 fs to be used in the presence of light hydrogen atoms in the system. In the above steps, turning off electrostatic interactions and employing a series of ad-hoc simulation procedures were implemented to eliminate close contacts between atoms, which is critical to avoiding the system to blow up in the subsequent simulation steps. We call the above-mentioned simulation procedures as the initialization process. We note that one can vary the simulation parameters in the initialization process as long as the resulted polymer structure does not lead to failure in the following annealing simulation.

The obtained polymer system was then annealed with electrostatic interactions turned on, and the PPPM (Particle–Particle–Particle–Mesh)-based Ewald sum method was used to calculate the electrostatic interactions. For the Lennard-Jones interactions, a cutoff of 0.800 nm was used. In the annealing process, the system was first simulated in an NPT ensemble at 1 atm and 1000 K (i.e., the annealing temperature) for 2 ps with timestep 0.1 fs applied. After that, the system was cooled from 1000 K to 300 K with a cooling rate of 140 K/ns in an NPT ensemble at 1 atm (referred to as the annealing process) followed by another NPT run at 300 K and 1 atm for 8 ns (referred to as the relaxation process) to achieve the final amorphous state, with a timestep of 1 fs and SHAKE constraints[33] applied. During annealing, the simulation box size was allowed to relax but constrained to have a cubic shape. These procedures are not designed to mimic experimental annealing but are used as a means to reliably achieve amorphous polymer states. Similar methods have been used extensively for other polymer studies.[17, 34] Such a pipeline enables high-throughput amorphous polymer structure generation with little to no human curation. Simulation conditions (e.g., annealing temperature and time) are also varied to examine how they would influence the calculated TC. An example optimization input LAMMPS script can be found in the supplemental information (SI).

**TC calculation**. Each relaxed amorphous polymer structure, which was in a cubic box, was then duplicated in three copies to form a cuboid (see **Figure 1a**). Depending on the specific polymers simulated, the dimensions varied a little due to density differences, but they were around 9.900 × 3.300 × 3.300 nm$^3$. The box was then used to calculate the TC via NEMD simulations. In the NEMD simulation, the system ran in an NVE (constant number of atoms, volume, and energy) ensemble for 5 ns with a timestep of 0.25 fs. The timestep of 0.25 fs was used to capture the vibration dynamics of the light hydrogen atoms. Larger timesteps would lead to energy drifting when the system includes hydrogen atoms. In these simulations, no SHAKE constraints were used since they would interfere the natural vibration of atoms, which is key to thermal transport. Heat source (320 K) and sink (280 K) were imposed using Langevin thermostats near the ends of the system (**Figure 1a**). The thickness of the source and sink regions were 0.500 nm. A fixed region at each end was implemented. The recorded heat flux, which was calculated from the heat addition/subtraction from the Langevin heat baths (**Figure 1b**), and the temperature profile (**Figure 1c**) were averaged over the last 4 ns of the production run, and they were used to calculate the TC via Fourier's law,

$$\kappa = -\frac{J}{\nabla T} \quad (1)$$

where $\kappa$ is TC, $J$ is heat flux, $\nabla T$ is temperature gradient obtained from the linear fit of the temperature profile. In the production run, the data of the last 4 ns was divided into 8 time periods with each producing a TC value. The standard deviations of these values were calculated (Table S1 in the SI).

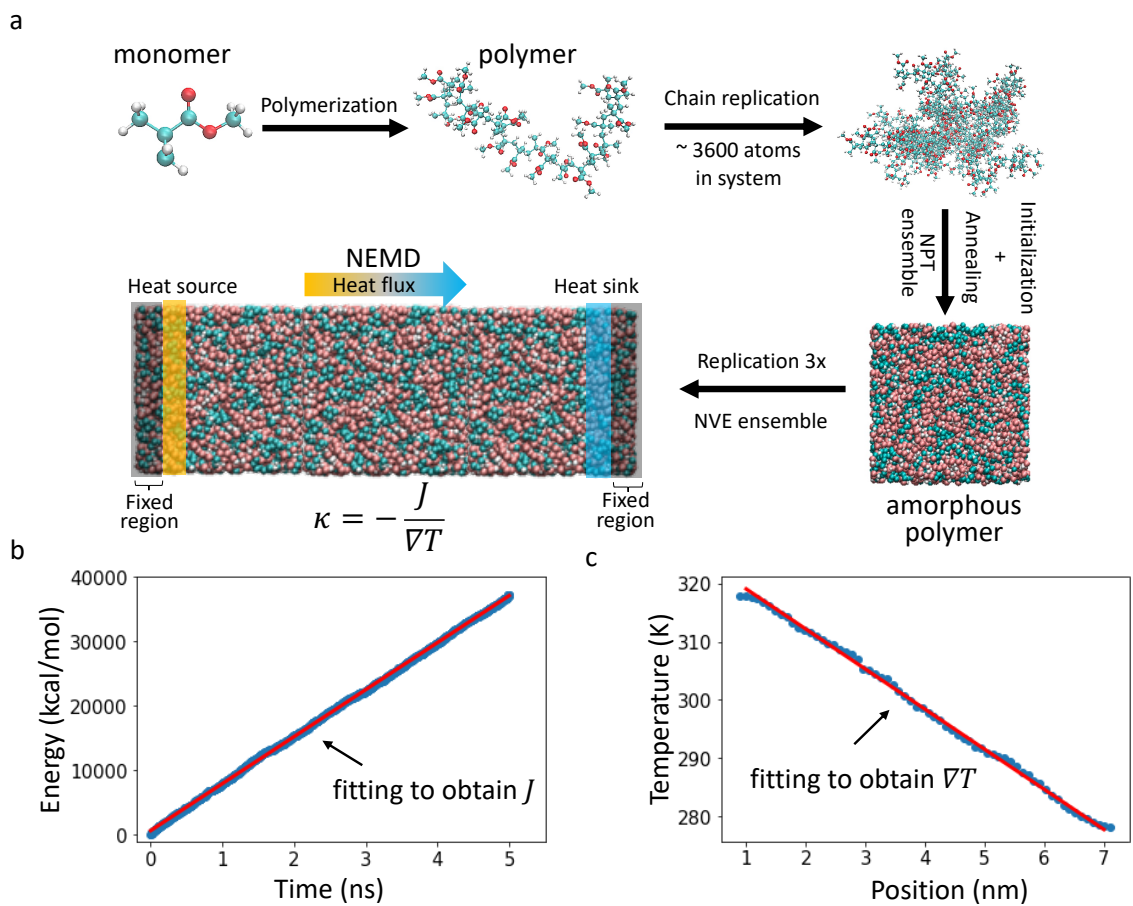

**Figure 1**. (a) Diagram of amorphous polymer generation and TC calculation using MD simulations: the amorphous polymer generation includes the polymerization of monomers, chain replication, fully structural relaxation, and TC calculation using NEMD via Fourier's law; (b) an example of energy added to or subtracted from the thermostated regions (i.e., heat source and sink) in the NEMD simulation; (c) an example steady-state temperature profile of the polymer system in the NEMD simulation.

**RESULTS**

**MD-Calculated TC.** We first randomly selected 365 polymers out of the 12,777 in the PoLyInfo database to calculate TC. We also took 62 polymers that have experimental TC in the PoLyInfo database for comparison, and our MD-labeled TC agreed generally in trend with the available experimental data in PoLyInfo (**Figure 2**). However, it is obvious that the agreement is not perfect. One potential cause of the observed discrepancy comes from systematic bias in the experimental data. For the experimental TC from PoLyInfo, if there were multiple records for one polymer, we took the median value of them for comparison, but some polymer had large variation in the TC data. For example, the TC of poly(2-methylprop-1-ene) in the PoLyInfo database varies from 0.115 to 0.171 W/m-K under different experimental conditions with the measurement methods varying from transient hot-wire method to line-source method. Likewise, the recorded TC of poly(vinyl alcohol) varies from 0.001 to 0.310 W/m-K where the measurement methods vary from FTIR, photoacoustic, Thermowave analyzer to hot-disk thermal constants analyzer. This is not unique to the experimental data in PoLyInfo. For example, the three nylon polymers (nylon 46, nylon 66, nylon 610) are reported to have different TC when measured using different methods.[35] In addition, MD-calculated TC can also have uncertainty related to the intrinsic stochastic nature of the amorphous polymer structure[4] and the lack of diversity in a single MD simulation. We have thus performed extensive test on the impact of simulation conditions, with the focus on annealing temperature and cooling rate, on the MD-calculated TC.

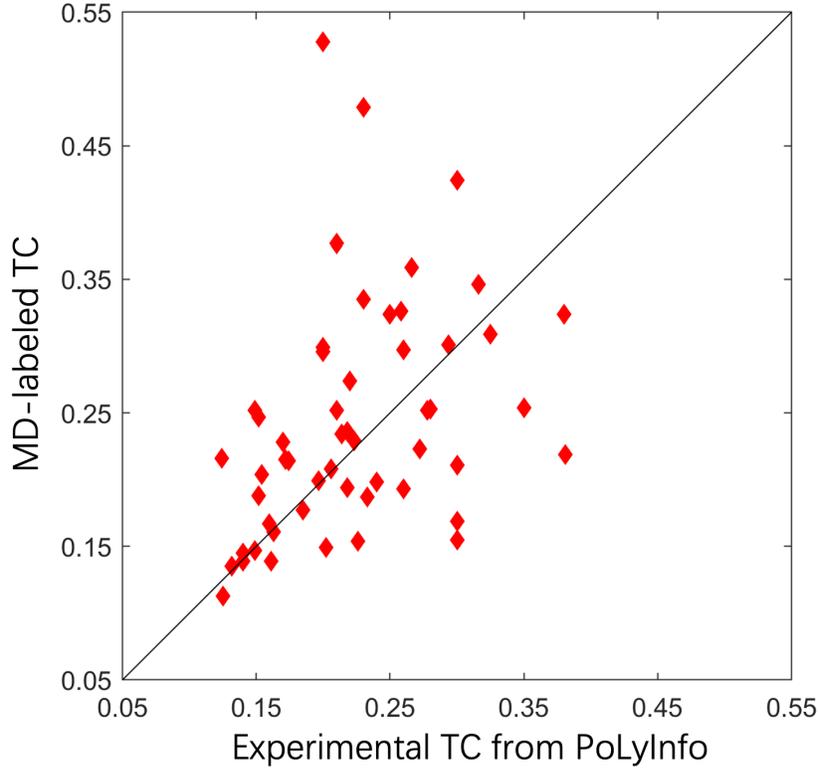

**Figure 2**. Pair plot between experimentally measured TC (from PoLyInfo) and MD-labeled TC.

We choose a number of polymers with different MD-labeled TC values and used several different simulation conditions to re-calculate their TC (**Figure 3b-d**). In one set of tests, we used different annealing temperatures, $T_a$, while the simulation settings in other processes (initialization, relaxation, and TC calculation processes) and cooling rate (100 K/ns) were fixed (**Figure 3a**). Three new annealing temperatures, 600, 1500 and 2000 K have been used, and 15 polymers were tested. Polymers selected for these tests had a wide range of MD-labeled TC (0.172 – 0.394 W/m-K). The results are shown **Figure 3b-d**, which show the parity plots between the TC from the simulations with the additional annealing temperatures and that from the original calculations which have $T_a$ = 1000 K. The variation in TC, characterized by the standard deviation of data shown in **Table 1**, can be up to 0.040 W/m-K, and such variation is generally larger for those polymers with higher TC, as also can be seen from **Figure 3b-d**. However, there does not seem to be a systematic change of TC with respect to the $T_a$, i.e., the variation is random. Thus, it is not conclusive that one $T_a$ can yield more accurate results than another. However, there were overall positive correlation between TC data from simulations with different $T_a$ values. The numerical data are shown in **Table 1**.

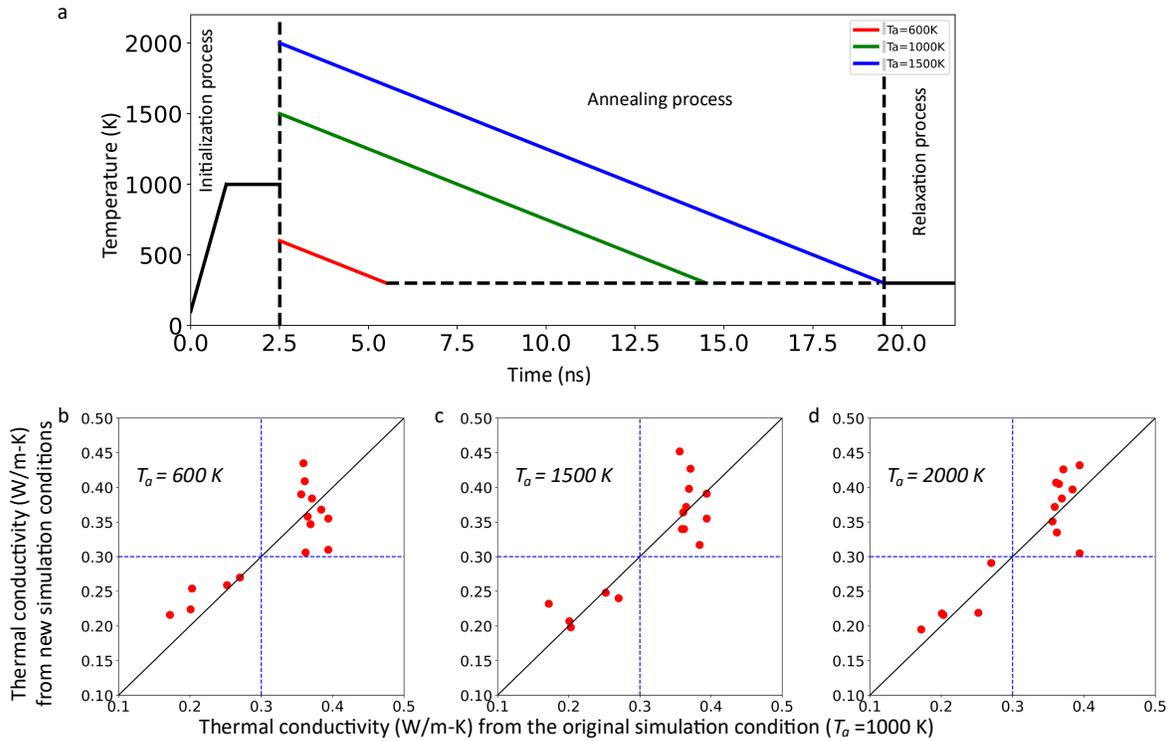

**Figure 3**. (a) Illustration of different simulation conditions with different annealing temperatures ($T_a$); (b-d) MD-labeled TC in new testing simulation conditions with different $T_a$ (600, 1500 and 2000 K) vs. MD-labeled TC in the original simulation condition ($T_a$ = 1000 K). The x-axis is for the original simulation condition, and the y-axes are for new testing simulation conditions.

**Table 1**. MD-labeled TC in new testing simulation conditions with different $T_a$ (600, 1500 and 2000 K) vs. MD-labeled TC in the original simulation condition ($T_a$ = 1000 K). All data units are W/m-K. Data is plotted in **Figure 3**.

| PID | Name | $T_a$ = 1000 K (original) | $T_a$ = 600 K | $T_a$ = 1500 K | $T_a$ = 2000 K | mean | STD |
|---|---|---|---|---|---|---|---|
| P010002 | Poly(prop-1-ene) | 0.252 | 0.259 | 0.248 | 0.219 | 0.245 | 0.020 |
| P010014 | Poly(2-methylprop-1-ene) | 0.201 | 0.224 | 0.207 | 0.218 | 0.213 | 0.018 |
| P010021 | Poly(vinylcyclopropane) | 0.172 | 0.216 | 0.232 | 0.195 | 0.204 | 0.019 |
| P010026 | Poly(allylcyclohexane) | 0.203 | 0.254 | 0.198 | 0.216 | 0.218 | 0.025 |
| P010042 | Poly(dodec-1-ene) | 0.362 | 0.306 | 0.340 | 0.335 | 0.336 | 0.036 |
| P010044 | Poly(octadec-1-ene) | 0.365 | 0.358 | 0.372 | 0.405 | 0.375 | 0.027 |
| P010077 | Poly(cyclodecene) | 0.369 | 0.347 | 0.398 | 0.384 | 0.375 | 0.036 |
| P010078 | Poly(docos-1-ene) | 0.371 | 0.384 | 0.427 | 0.426 | 0.402 | 0.040 |
| P010080 | Poly(hexadec-1-ene) | 0.394 | 0.310 | 0.355 | 0.305 | 0.341 | 0.030 |
| P010098 | Polycyclopentene | 0.270 | 0.270 | 0.240 | 0.291 | 0.268 | 0.023 |
| P010106 | Poly(cyclododecene) | 0.394 | 0.355 | 0.391 | 0.432 | 0.393 | 0.015 |
| P020056 | Poly(4-octadecylstyrene) | 0.359 | 0.435 | 0.340 | 0.372 | 0.364 | 0.009 |
| P034502 | Poly(octadecyl vinyl ether) | 0.356 | 0.390 | 0.452 | 0.351 | 0.387 | 0.023 |
| P040135 | Poly(N-docosylacrylamide) | 0.384 | 0.368 | 0.317 | 0.397 | 0.367 | 0.022 |
| P070101 | Poly(pentamethyleneoxydecamethyleneoxydecamethylene oxide) | 0.361 | 0.409 | 0.364 | 0.407 | 0.385 | 0.018 |

One potential influencer of the TC variation is density.[4] The densities of those optimized polymer structures were recorded in their TC calculation processes, for both the original simulation condition and the three new simulation conditions. As shown in **Table S2** of SI, the variation in density was small in different simulation conditions and did not correspond to the variation in the TC data, suggesting that the variation in TC is likely from the randomness of morphology from different simulations. In addition, we have tested a number of other polymers with diverse TC and compared their standard deviation within one simulation and that from five ensembles (Table S1 and Figure S1 in SI). It was found that within each NEMD simulation, the TC uncertainty is more consistent for all the tested polymer (7-10%), but the ensembles have more spread percentile uncertainty (7-20%). The disparity between these two uncertainties increased as TC increased. This finding also suggests that it is the randomness in morphology that leads to TC variation, as rigid polymers usually have larger TC[4] and are usually more difficult to reach global minimal structure and thus more structural diversity in MD simulations. As shown later, high TC polymers indeed turn out to be more rigid polymers.

Besides the annealing temperature, the cooling rate in the annealing process was also tested to examine their impact on the MD-labeled TC. Thus, we also re-calculated the TC for 6 polymers with different annealing rates ($\tau_a$) but with the same $T_a$ = 1500 K. These polymers were selected based on their diversity in the MD-labeled TC from the original calculations. These additionally tested $\tau_a$, besides the original 140 K/ns, were 80, 100, 120 and 150 K/ns (**Figure 4a**), respectively. The simulation settings in other processes (initialization, relaxation, and TC calculation processes) and annealing temperature (1500 K) were fixed. From **Figure 4b-e**, the different annealing rate led to less variations in the predicted TC, and there was again no systematic trend of TC dependency on the annealing rate observed. Still, those polymers with TC > 0.300 W/m-K from the original calculation are all above this threshold in the new test simulation conditions. The data used for **Figure 4** are shown in **Table 2**.

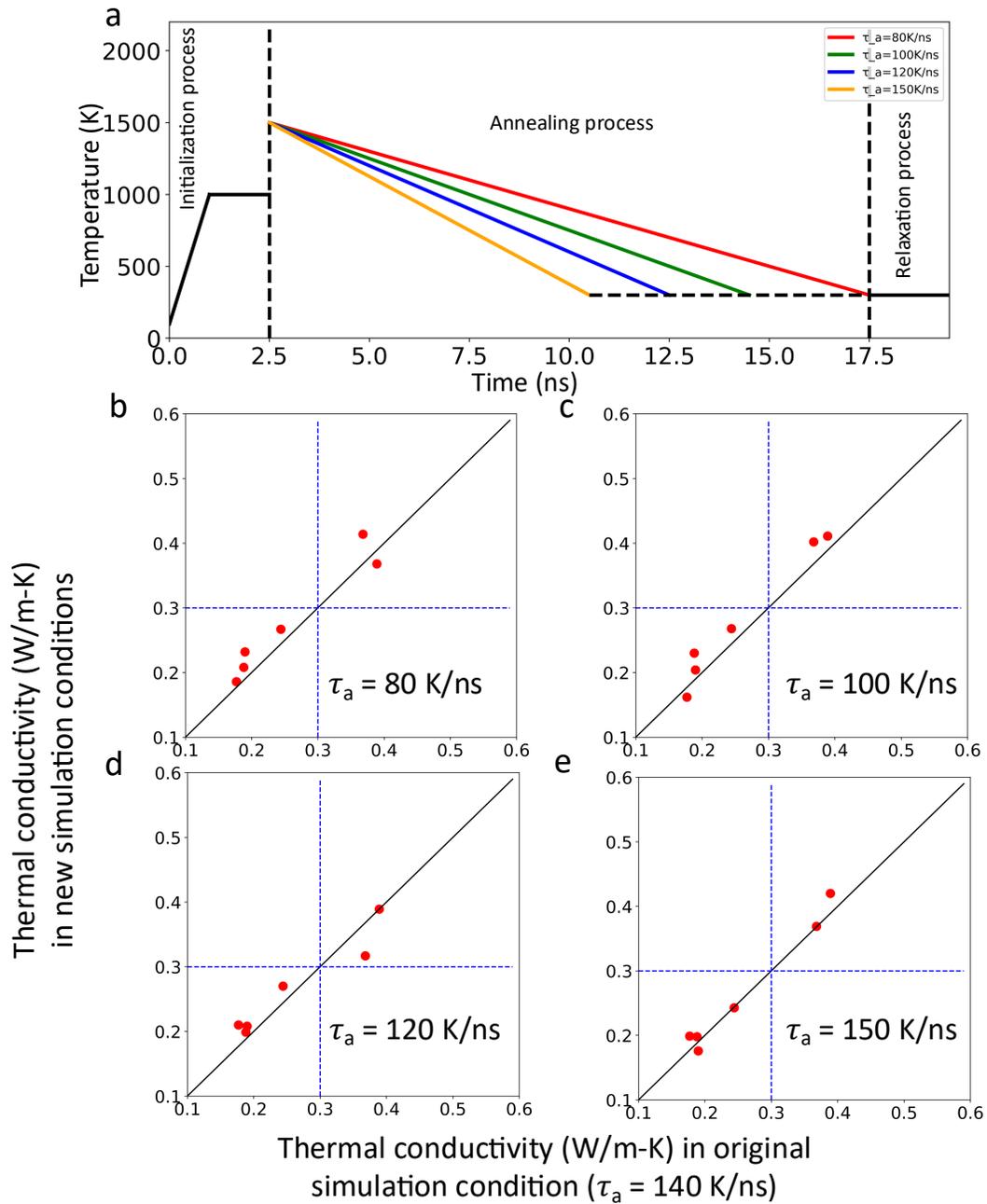

**Figure 4**. (a) Illustration of simulation conditions with different annealing rates ($\tau_a$); (b-e) MD-labeled TC in new simulation conditions vs. MD-labeled TC in the original simulation condition. The TC at the x-axis corresponds to the those calculated in the original simulation condition, and the TC at the y-axes corresponds to those calculated in the new simulation conditions.

**Table 2**. MD-labeled TC in new simulation conditions with different $\tau_a$ (80, 100, 120, 150 K/ns) vs. MD-labeled TC from the original simulation condition ($\tau_a$ =140 K/ns). All data units are W/m-K. Data is plotted in **Figure 4**.

| PID | Name | $\tau_a$ =140 K/ns (original) | $\tau_a$ ($T_a$ = 1500 K) | | | | mean | STD |
|---|---|---|---|---|---|---|---|---|
| | | | 80 K/ns | 100 K/ns | 120 K/ns | 150 K/ns | | |
| P010009 | Poly(hept-1-ene) | 0.244 | 0.267 | 0.268 | 0.270 | 0.243 | 0.258 | 0.012 |
| P010052 | Poly(4-methylhex-1-ene) | 0.190 | 0.232 | 0.204 | 0.208 | 0.176 | 0.202 | 0.019 |
| P010006 | Poly(3-methylbut-1-ene) | 0.177 | 0.186 | 0.162 | 0.210 | 0.199 | 0.187 | 0.017 |
| P020001 | Polystyrene | 0.188 | 0.208 | 0.230 | 0.199 | 0.198 | 0.205 | 0.014 |
| P090099 | Poly(hexadecanolactone) | 0.389 | 0.368 | 0.411 | 0.389 | 0.420 | 0.395 | 0.018 |
| P100233 | Poly[(dodecane-1,12-diamine)-alt-(decanedioic acid)] | 0.368 | 0.414 | 0.402 | 0.317 | 0.369 | 0.374 | 0.034 |

**Regression model.** According to these tests, it is understood that the TC data generated from MD simulations have uncertainties, but such uncertainties do no prevent one from using the simulations to predict a trend in TC or tell if a polymer is likely to have relatively high TC (> 0.300 W/m-K) or low TC. As a result, we used the generated 365 TC data from the original simulation conditions to perform two machine learning tasks, including a regression task and a classification task.

In the regression task, we established a surrogate model for the polymer chemistry-TC relation. Each polymer was first represented using the polymer embedding (a continuous-valued vector in the length of 300), which is a machine-learned polymer representation trained using a combination of PoLyInfo and PI1M databases, totaling about 1,008,576 polymer structures.[36] The polymer embedding representation was obtained by training the mol2vec[37] model on the monomer structures. In mol2vec, monomers were first decomposed into sequence of substructures, where the center substructure was used as input of a single-hidden-layer neural network to predict its surrounding substructures. Each substructure in the monomer was used as center substructure for once and the polymer embedding was derived from the weights of the hidden layer after training was done. More details of polymer embedding can be found in Ref. [36]. PI1M is a virtual library of polymer SMILES sampled from a recurrent neural network (RNN) model trained based on the PoLyInfo polymer structures. We then trained an Random Forests (RF) model [38] to quantify the relationship between the polymer embedding and the MD-labeled TC. The number of trees in the RF was set to 1,500 based on grid search in hyperparameter optimization. The RF was trained in the manner of 10-fold cross-validation, and we call this trained model RF-R (i.e., random forests model for regression). The predictive accuracy, estimated by the $R^2$, is 0.789±0.030 on validation sets, and the pair plot between RF-R-predicted TC and MD-calculated TC is shown in **Figure 5a**. The distribution of MD-labeled TC is shown in **Figure 5b**, the range of which is between 0.035 and 0.395 W/m-K, with the majority below 0.300 W/m-K (324 out of 365, or 89%). We believe that there should be more polymers with TC greater than 0.300 W/m-K in the PoLyInfo database.

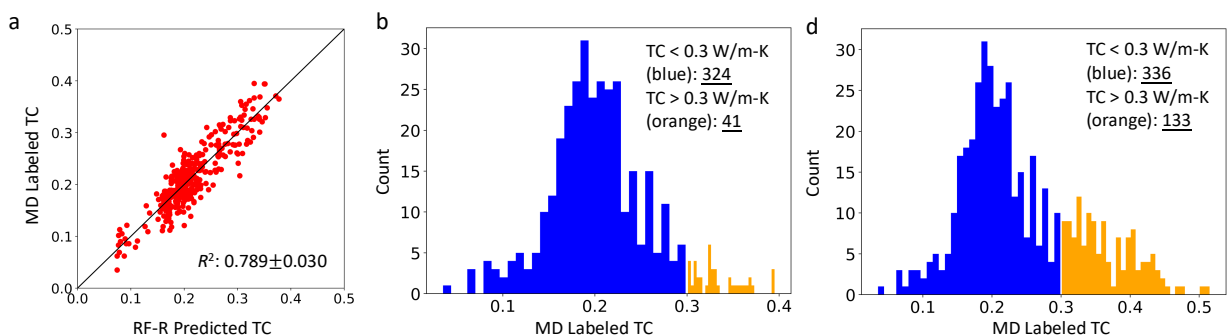

**Figure 5.** (a) Pair plot between RF-R-predicted TC and MD-labeled TC on the validation sets of 365 training data; (b) distribution of the initial 365 MD-labeled TC, with 324 below 0.300 W/m-

K and 41 above 0.300 W/m-K; (c) distribution of the total 469 MD-labeled TC, with 336 below 0.300 W/m-K and 133 above 0.300 W/m-K.

Rather than continuing to generate data randomly in the chemical space, we used RF-R to exploit polymers with relatively high TC. More specifically, we predicted the TC (i.e., pseudo label) for all 12,777 polymers in the PoLyInfo database using the RF-R model to identify polymer candidates that have RF-R-predicted TC greater than 0.300 W/m-K but have not been MD-labeled. We choose 0.300 W/m-K as the threshold because the tests in the previous sections indicate that the MD simulations tend to correctly predict polymers with TC above or below this threshold despite the calculation uncertainty. We found 880 polymers with RF-R-predicted TC greater than this threshold, and we randomly selected 104 of them to label using MD simulations. Among these 104 polymers, 92 of them (or 88%) had TC greater than 0.300 W/m-K in contrast to 41/365 (or 11%) in the original dataset. The TC distribution of the total 469 MD-labeled polymers is shown in **Figure 5c**, which shows a much higher population above 0.300 W/m-K.

In **Figure 6**, we visualize the top 35 polymers with the highest MD-labeled TC and list their corresponding polymer classes. These polymers are MD-labeled to have TC > 0.400 W/m-K. The majority of them belong to polyamides (10/35) and polysulfides (12/35), and the rest are polyvinyls, polyoxides, polyimines, polyolefins, and polyesters. The fact that polyamides are predicted to have high TC is consistent with a prior study,[28] where three thermally conductive polymers designed all belong to polyamides. Given the uncertainty in MD calculations as discussed above, to further confirm the high-TC nature of those 35 polymers, we re-simulated them with different initial structures using the original simulation condition, as we have concluded that morphology randomness is a major cause of the variation in MD-labeled TC. Each time the MD simulation pipeline was executed, it produces a new initial structure due the randomness in packing the molecule chains into the simulation box. We run the pipeline for another two times to obtain two new initial structures for these 35 polymers, and we calculated the TC using the original simulation procedures. The MD-labeled TC for those two different initial structures is shown in **Table 3**. When taking the average of MD-labeled TC values, as can be seen from **Table 3**, 22 out of 35 still had MD-labeled TC above 0.400 W/m-K, and all, except one, had TC above 0.360 W/m-K.

**Table 3**. MD-labeled TC for the 35 high-TC polymers with different initial structures. All data units are W/m-K.

| PID | Name | Original | Different initial structure 1 | Different initial structure 2 | Mean | STD |
| --- | --- | --- | --- | --- | --- | --- |
| P070069 | Poly(dodecamethylene oxide) | 0.473 | 0.419 | 0.400 | 0.431 | 0.031 |
| P070096 | Poly(methyleneoxyoctadecamethylene oxide) | 0.431 | 0.372 | 0.415 | 0.406 | 0.025 |
| P070099 | Poly(decamethyleneoxyhexamethyleneoxyhexamethylene oxide) | 0.416 | 0.325 | 0.344 | 0.362 | 0.039 |
| P402250 | Poly[(undecane-1,11-diamine)-alt-(icosanedioic acid)] | 0.406 | 0.317 | 0.392 | 0.372 | 0.039 |
| P100018 | Poly[(octane-1,8-diamine)-alt-(docosanedioic acid)] | 0.432 | 0.448 | 0.432 | 0.437 | 0.008 |
| P100029 | Poly[(dodecane-1,12-diamine)-alt-(octadecanedioic acid)] | 0.404 | 0.355 | 0.483 | 0.414 | 0.053 |
| P100229 | Poly[(decane-1,10-diamine)-alt-(docosanedioic acid)] | 0.424 | 0.333 | 0.408 | 0.388 | 0.040 |
| P100235 | Poly[(dodecane-1,12-diamine)-alt-(dodecanedioic acid)] | 0.423 | 0.477 | 0.376 | 0.425 | 0.041 |
| P100238 | Poly[(tetradecane-1,14-diamine)-alt-(docosanedioic acid)] | 0.456 | 0.389 | 0.435 | 0.427 | 0.028 |
| P100240 | Poly[(octadecane-1,18-diamine)-alt-(dodecanedioic acid)] | 0.515 | 0.415 | 0.371 | 0.434 | 0.060 |
| P100696 | Poly(22-aminodocosanoic acid) | 0.417 | 0.531 | 0.383 | 0.444 | 0.063 |
| P312010 | Poly[(buta-1,3-diene)-alt-(1-dodecene)] | 0.442 | 0.375 | 0.401 | 0.406 | 0.028 |
| P312012 | Poly[(buta-1,3-diene)-alt-(hexadec-1-ene)] | 0.402 | 0.453 | 0.354 | 0.403 | 0.040 |
| P332076 | Poly(22-tricosynoic acid) | 0.435 | 0.357 | 0.417 | 0.403 | 0.033 |
| P332081 | Poly(22,24-pentacosadiynoic acid) | 0.402 | 0.354 | 0.447 | 0.401 | 0.038 |
| P332304 | Poly[1-(2,5,8,11,14-pentaoxaicosan-1-yl)henicosane-1,21-diyl] | 0.423 | 0.349 | 0.426 | 0.399 | 0.036 |
| P332306 | Poly[1-(2,5,8,11,14-pentaoxaoctacosan-1-yl)henicosane-1,21-diyl] | 0.433 | 0.375 | 0.345 | 0.384 | 0.037 |
| P373742 | Poly(2-hexadecylthieno[3,4-b][1,4]dioxane-5,7-diyl) | 0.506 | 0.391 | 0.437 | 0.445 | 0.047 |
| P380064 | Poly(tetradecamethylene sulfide) | 0.402 | 0.309 | 0.313 | 0.341 | 0.043 |
| P380068 | Poly(3-tetradecylthiophene) | 0.439 | 0.468 | 0.348 | 0.418 | 0.051 |
| P380123 | Poly{(2,5-disulfanyl-p-phenylenediamine)-alt-[2,5-bis(dodecyloxy)terephthaloyl dichloride]} | 0.404 | 0.336 | 0.354 | 0.365 | 0.029 |
| P382002 | Poly(3-tridecylthiophene) | 0.468 | 0.415 | 0.331 | 0.405 | 0.056 |
| P382003 | Poly(3-pentadecylthiophene) | 0.413 | 0.406 | 0.304 | 0.374 | 0.050 |
| P382013 | Poly{[2,5-dibromo-3,4-bis-(dodecyloxy)thiophene]-alt-[1,2-bis(tributylstannio)ethene]} | 0.418 | 0.328 | 0.349 | 0.365 | 0.038 |
| P382020 | Poly(3-docosylthiophene) | 0.428 | 0.327 | 0.437 | 0.397 | 0.050 |
| P382079 | Poly[1,4-bis(3-dodecyl-2-thienyl)bezene] | 0.400 | 0.418 | 0.421 | 0.413 | 0.009 |
| P382158 | Poly(3-octadecylthiophene) | 0.442 | 0.340 | 0.314 | 0.365 | 0.055 |
| P382279 | Poly(3-(dodecyloxy-4-methylthiophene)) | 0.430 | 0.377 | 0.329 | 0.379 | 0.041 |
| P382280 | Poly[3-methyl-4-(tetradecyloxy)thiophene] | 0.408 | 0.456 | 0.446 | 0.437 | 0.021 |
| P382282 | Poly[3-(icosyloxy)-4-methylthiophene] | 0.410 | 0.399 | 0.315 | 0.375 | 0.042 |
| P392423 | Poly[(butane-1,4-diol)-alt-(hexatriacontanedioic acid)] | 0.421 | 0.513 | 0.550 | 0.495 | 0.054 |
| P402128 | Poly(iminododecane-1,12-diyliminoicosane-1,20-dioyl) | 0.449 | 0.443 | 0.428 | 0.440 | 0.009 |
| P402250 | Poly[(undecane-1,11-diamine)-alt-(icosanedioic acid)] | 0.406 | 0.429 | 0.447 | 0.427 | 0.017 |
| P402252 | Poly[(heptane-1,7-diamine)-alt-(icosanedioic acid)] | 0.414 | 0.351 | 0.476 | 0.414 | 0.051 |
| P462521 | Poly[(dimethylazaniumdiyl)hexane-1,6-diyl(dimethylazaniumdiyl)hexadecane-1,16-diyl] | 0.445 | 0.367 | 0.412 | 0.408 | 0.032 |

**Polyamides**

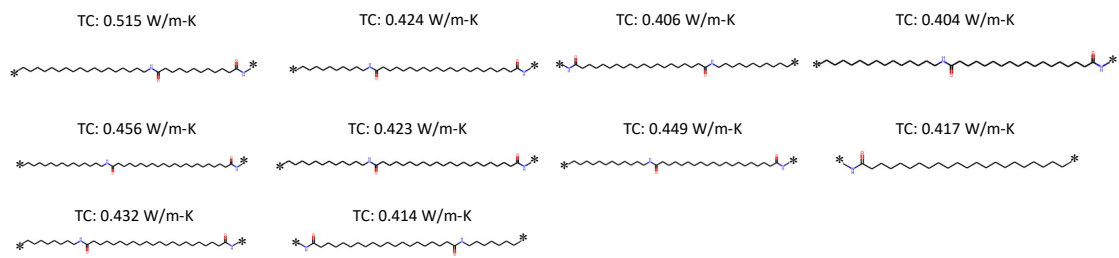

**Polysulfides**

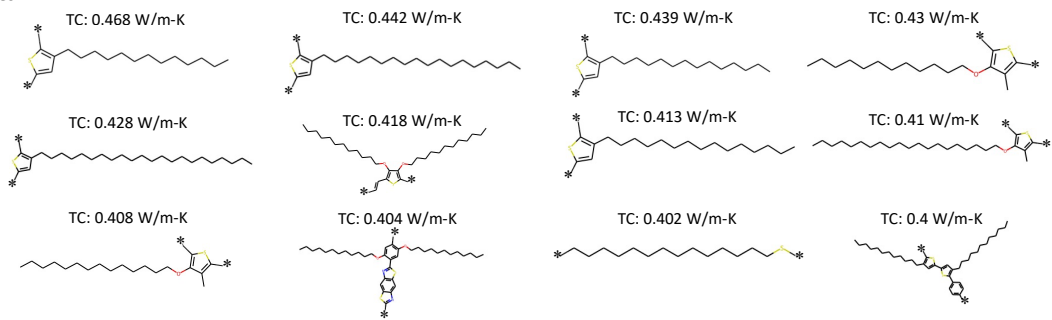

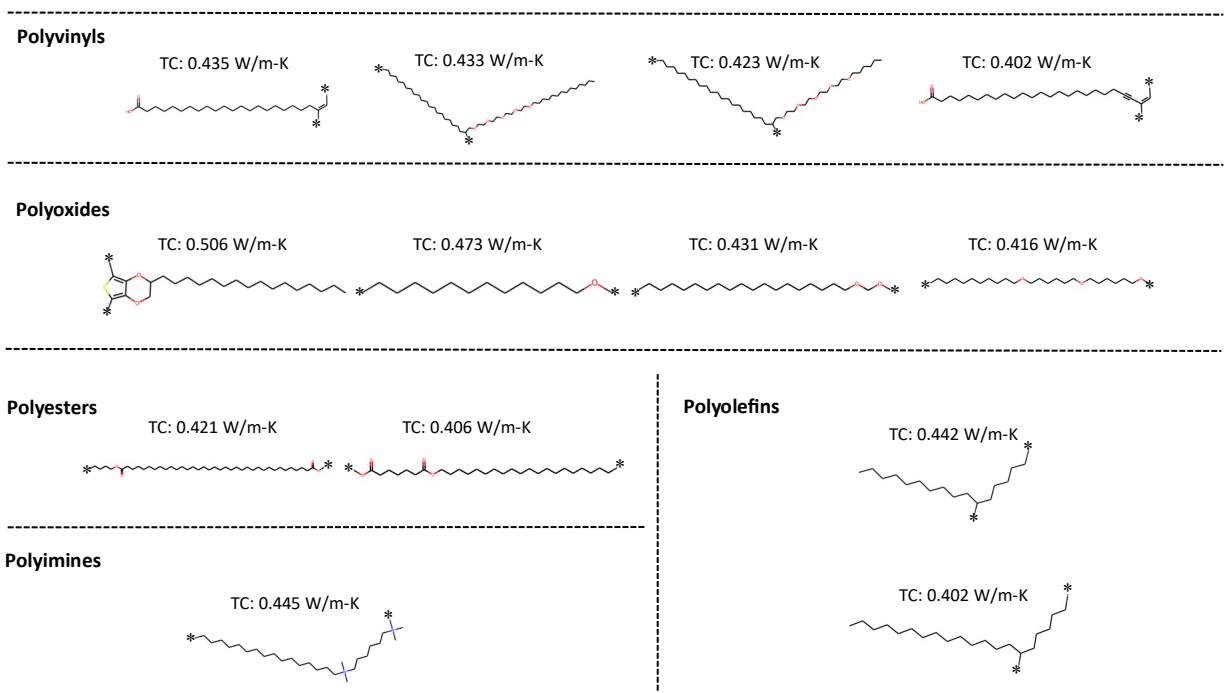

**Figure 6**. The polymer structures with MD-labeled TC above 0.400 W/m-K, with their corresponding polymer classes indicated. Red, blue, and yellow colors indicate oxygen, nitrogen, and sulfur atoms respectively.

As can be seen from **Figure 5c**, the highest MD-labeled TC identified in the original simulation condition was from nylon 18,12 with a value of 0.515 W/m-K. nylon 18,12 belongs to a type of polyamide polymer. Given the known uncertainty of the MD simulations as discussed above, we scrutinized the TC of this polymer further. We have performed an additional ten simulations with different simulation conditions as shown in **Table 4**. From cases 1 to 3, we only changed the duration of relaxation process at 300 K, but observed little variation in MD-labeled TC, suggesting that MD-labeled TC is not sensitive to the relaxation at 300 K. This is understandable since at such a low temperature, the mobility of the polymer chain should be very small, which cannot cause significant change in morphology and then TC. In cases 4 and 5, we changed both the annealing temperature and annealing rate while keep the relaxation at 300 K the same. We observed an obvious decrease in TC compared to the first three cases. The limited change in the pre-heating time (cases 6 and 7) did not lead to significant change in TC. We note that longer pre-heating processes tend to lead to crashes in simulations. It is our belief that compared to the long annealing process when the polymer system can stay under high temperature for a sufficient amount of time (e.g., the polymer system can stay above 1000K for 4.2 ns if the system is annealed from 1500K to 300K in 10 ns), the short pre-heating process seems to be unnecessary. In cases 8 to 11, we removed the pre-heating process and further changed the annealing temperature but with a constant annealing rate of 100 K/ns. These changes in annealing temperature, however, did not provide a clear trend in TC. Overall, depending on the simulation conditions, the TC of nylon 18,12 varied with a mean of 0.429 W/m-K and a standard deviation of 0.064 W/m-K, which still place it in the high end of TC. It is interesting to find that these averaged value and standard deviation are similar to those in Table 3 for this polymer (0.434 W/m-K, 0.060 W/m-K) which used different initial structures for ensemble averaging. This also suggest that different simulation procedures provide diverse random amorphous structures that lead to the variation in TC.

**Table 4**. MD-calculated TC for nylon 18, 12 under different simulation conditions.

| Case | Pre-heating (NPT) | | | Annealing (NPT) | | | | relaxation at 300K (NPT) | TC |
|---|---|---|---|---|---|---|---|---|---|
| | 1000K | 1500K | 2000K | 1000K to 300K | 1500K to 300K | 1700K to 300K | 2000K to 300K | | |
| 1 | 2 ps | - | - | 5 ns | - | - | - | 8 ns | 0.515 W/m-K |
| 2 | 2 ps | - | - | 5 ns | - | - | - | 12 ns | 0.524 W/m-K |
| 3 | 2 ps | - | - | 5 ns | - | - | - | 20 ns | 0.530 W/m-K |
| 4 | - | 2 ps | - | - | 8 ns | - | - | 12 ns | 0.403 W/m-K |
| 5 | - | - | 2 ps | - | - | - | 10ns | 12 ns | 0.442 W/m-K |
| 6 | - | 10 ps | - | - | 10 ns | - | - | 2 ns | 0.377 W/m-K |
| 7 | - | 20 ps | - | - | 10 ns | - | - | 2 ns | 0.375 W/m-K |
| 8 | - | - | - | 7 ns | - | - | - | 2 ns | 0.366 W/m-K |
| 9 | - | - | - | - | 12ns | - | - | 2 ns | 0.387 W/m-K |
| 10 | - | - | - | - | - | 14 ns | - | 2 ns | 0.384 W/m-K |
| 11 | - | - | - | - | - | - | 17 ns | 2 ns | 0.423 W/m-K |

We have also compared our MD-calculated TC with available experimental data for a few other polyamide polymers (see **Table 5**). Based on the results listed in **Table 5** our MD-labeled TC values agree reasonably with experimental TC, except for nylon 6,10, which shows diverse experimental values from different experimental conditions. The TC of nylon 6,10 reported in ref. [35] was measured using a flash DSC (differential scanning calorimetry) method, while that reported in PoLyInfo was measured in a quasi-stationary method.

**Table 5**. Comparison between MD-labeled TC and experimentally measured TC for several polyamide polymers.

| Polymer Name | Experimental TC (W/m-K) | MD-labeled TC (W/m-K) |
| --- | --- | --- |
| Nylon 6, 6 | 0.250-0.270 [35, 39] | 0.252 |
| Nylon 4, 6 | 0.270-0.300 [35, 40] | 0.310 |
| Nylon 6, 10 | 0.230 [35], 0.325 (PoLyInfo) | 0.383 |
| Nylon 6 | 0.300 (PoLyInfo) | 0.298 |
| Nylon 12 | 0.316 (PoLyInfo) | 0.346 |

**Classification.** In the second machine learning task, the initial MD-labeled 365 polymers were used to train another RF model (with 1500 trees) to classify whether a polymer will have a TC above or below 0.300 W/m-K. We refer to this model as RF-C (i.e., random forests model for classification). The predictive accuracy of RF-C, evaluated by the ratio of correctly classified data and measured in 10-fold cross-validation, is 0.934±0.054. The confusion matrix on the validation set is shown in **Figure 7a**, indicating that most of the data (341/365) are correctly classified. The false omission rate is 2.500%, but the positive predictive value is 61%, mostly due to the imbalanced data (e.g., less data > 0.300 W/m-K). The receiver operating characteristics (ROC) curve on the validation data is shown in **Figure 7b**, which also indicate an accurate classification model, as the curve is skewed to the upper left corner with under the curve area of 0.970. Using the trained RF-C, 12,777 polymers from PoLyInfo were then classified, and 826 polymers that have not been simulated at the beginning were classified to have TC > 0.300 W/m-K. This is similar to the number (880) predicted using the RF-R (RF regressor) above, and there were 713 overlapping polymers proposed by both RF-R and RF-C. Out of the 104 randomly selected polymers with pseudo label > 0.300 W/m-K in the above regression task, 100 were predicted to be in this class (TC > 0.300 W/m-K) by RF-C, and 89 of them (or 89%) were proved to have MD-labeled TC > 0.300 W/m-K. This ratio of 89% is similar to that found in the regression task (88%).

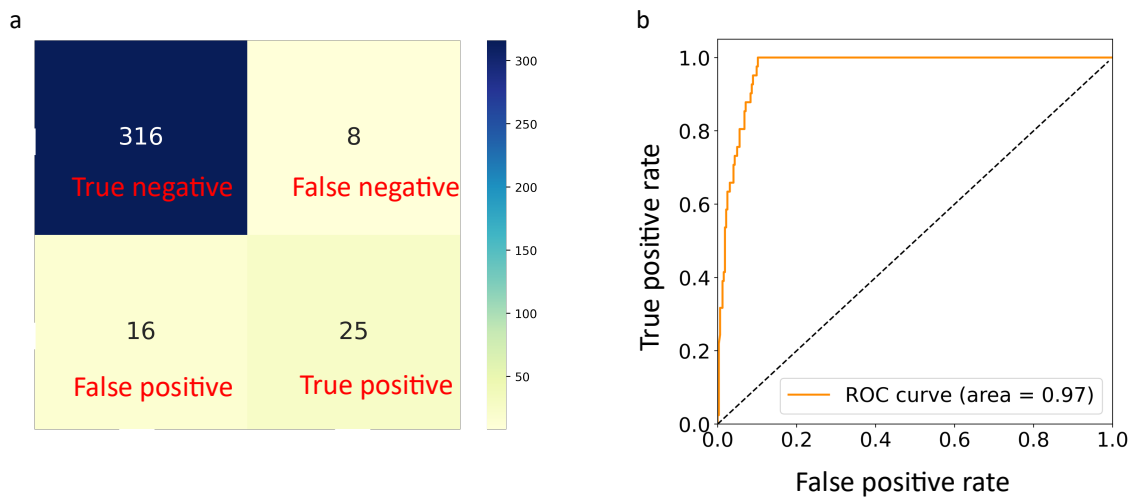

**Figure 7**. (a) The confusion matrix of the RF-C model, indicating the number of true negatives, false negatives, false positives, and true positives; (b) the ROC curve and the area under the curve.

**CONCLUSION**

In summary, we first used high-throughput MD simulations to calculate TC for a large number of polymers in the PoLyInfo database. Tests on annealing temperature and annealing rate in the simulation procedure were performed to examine their impacts on the calculated TC. The test results showed that the MD-calculated TC contains noise, but could still provide meaningful trends. For selected polymers, we also performed ensemble averaging studies. We found that within each NEMD simulation, the TC uncertainty was more consistent (7-10%), but the ensembles showed more spread percentile uncertainty (7-20%). The difference between these two uncertainties increased as TC increased. These findings suggested that it was the randomness in morphology that led to TC variation, as rigid polymers usually had larger TC and are usually more difficult to relax to the ground state and thus more structural diversity in MD simulations. The MD-generated TC data was then used to trained a random forests regression model, which was used to screen the polymers in the PoLyInfo database to identify candidates with TC > 0.300 W/m-K. Selected polymer candidates proposed by the regression model were further MD-labeled, and 133 polymers with TC > 0.300 W/m-K were eventually identified. Among them, 35 were found to have TC > 0.400 W/m-K, with majority of them belonging to polyamide and polysulfides classes, and the rest are polyvinyls, polyoxides, polyimines, polyolefins, and polyesters. The strategy and results from this work may provide useful guidance and an integral component to the experimental exploration of high TC polymers.

**ACKNOWLEDGEMENT**

The computation is supported in part by the University of Notre Dame, Center for Research Computing, and NSF through XSEDE resources provided by TACC Stampede II under a grant number TG-CTS100078.